\documentclass[conference,a4paper]{APSIPA2021}

\usepackage{multirow}
\usepackage{graphicx}
\usepackage{amsmath}
\usepackage[psamsfonts]{amssymb}
\usepackage{amsxtra}
\usepackage{threeparttable}

\title{End-to-End Mandarin Tone Classification with Short Term Context Information}

\author{%
  \authorblockN{%
    Jiyang Tang\authorrefmark{1} and
    Ming Li\authorrefmark{1}
  }
  \authorblockA{%
    \authorrefmark{1}
    Data Science Research Center, Duke Kunshan University, Kunshan, China \\
    E-mail: \{jiyang.tang, ming.li369\}@dukekunshan.edu.cn}
}

\begin{document}

  \maketitle
  \thispagestyle{empty} 

  \begin{abstract}
    In this paper, we propose an end-to-end Mandarin tone classification method from continuous speech utterances
    utilizing both the spectrogram and the short-term context information as the input.
    Both spectrograms and context segment features are used to train the tone classifier.
    We first divide the spectrogram frames into syllable segments using force alignment results produced by
    an ASR model.
    Then we extract the short-term segment features to capture the context information across multiple syllables.
    Feeding both the spectrogram and the short-term context segment features into an end-to-end model could
    significantly improve the performance.
    Experiments are performed on a large-scale open-source Mandarin speech dataset to evaluate the proposed method.
    Results show that this method improves the classification accuracy from $79.5\%$ to $92.6\%$ on the
    AISHELL3 database.
  \end{abstract}

  \section{Introduction}

  \subsection{Background}

  In Computer-Assisted Language Learning (CALL) systems for Mandarin learners, tone classification is one of
  the most important tasks.
  In Mandarin, there are four lexical tones, denoted as T1 to T4, plus a neutral tone, denoted as T5.
  Also, some syllables are considered not to have a tone, and they are denoted as T5.
  Mandarin tones have important lexical meanings.
  For example, \textit{shi1} (poem), \textit{shi2} (time), \textit{shi3} (let) and \textit{shi4} (matter) have
  T1, T2, T3, and T4 as their tones.
  They have completely different meanings, and their pronunciation is distinguished solely by the tones.

  In terms of pronunciation, T1 is a flat tone with a relatively high frequency, T2 has a rising frequency,
  T3 first dips and then rises, and T4's frequency falls from a relatively high value
  \cite{tone_nucleus,tone_classification_cnn,tone_recognition_conv_blstm_att,ToneNet,contextual_tonal_variations}.
  T5 itself doesn't possess a particular pitch contour pattern.
  Instead, it is a short and unstressed continuation of the previous syllable
  \cite{tone_nucleus,tone_recognition_nn,tone_modeling}.
  Because the first part of a bi-syllabic Chinese character, called the initial (shengmu),
  does not contribute much to the tone of a syllable, it is not considered to have a tone attached, denoted as T0.
  It is also referred to as ``onset course'' in the tone nucleus model \cite{tone_nucleus}.

  Continuous tone classification is a difficult task.
  The main issue is that the pitch contours of a syllable can vary drastically depending on the surrounding syllables.
  Because the human glottis cannot shift from one frequency to another instantaneously, there has to be a transition
  between two different tones in a continuous speech utterance.
  Thus, the pitch contours of a syllable can be affected by the context bi-directionally
  \cite{contextual_tonal_variations}.
  This phenomenon is called tonal coarticulation \cite{tone_recognition_conv_blstm_att,contextual_tonal_variations,
    tone_recognition_nn,tone_recognition_without_pitch,discrim_tonal_feats}.

\begin{figure}[h]
    \centering
    \includegraphics[width=60mm]{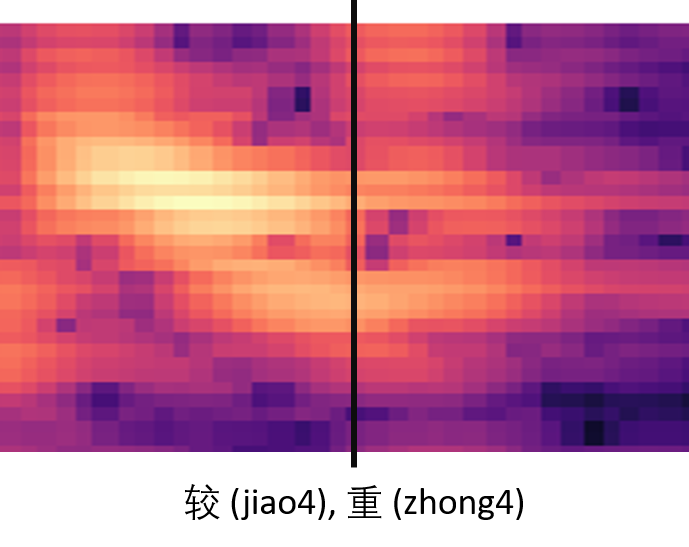}
    \caption{Example of two consecutive T4 segments, where the second one is very difficult to distinguish}
    \label{fig:difficult_example}
  \end{figure}

  Tonal coarticulation makes it very difficult to recognize the tone of a syllable directly using pitch contours.
  Figure \ref{fig:difficult_example} shows two consecutive T4 syllables.
  The first syllable ends with a relatively high frequency, while the second one starts with a low frequency.
  The transition between the two syllables makes the pitch contours very blurry around the syllable boundaries (the
  vertical black line).

  Moreover, ``tone sandhi'', which is a Mandarin pronunciation rule that changes the pronounced tones in certain
  situations, makes it more difficult for tone classification \cite{contextual_tonal_variations,
    tone_recognition_without_pitch}.
  For example, two consecutive T3 are usually pronounced as a T2 followed by a T3 in continuous speech.

  In this paper, we propose a continuous Mandarin tone classification framework that utilizes short-term context
  information to reduce the negative performance impact created by the tonal coarticulation.

  \begin{figure*}[t]
    \begin{center}
      \includegraphics[width=140mm]{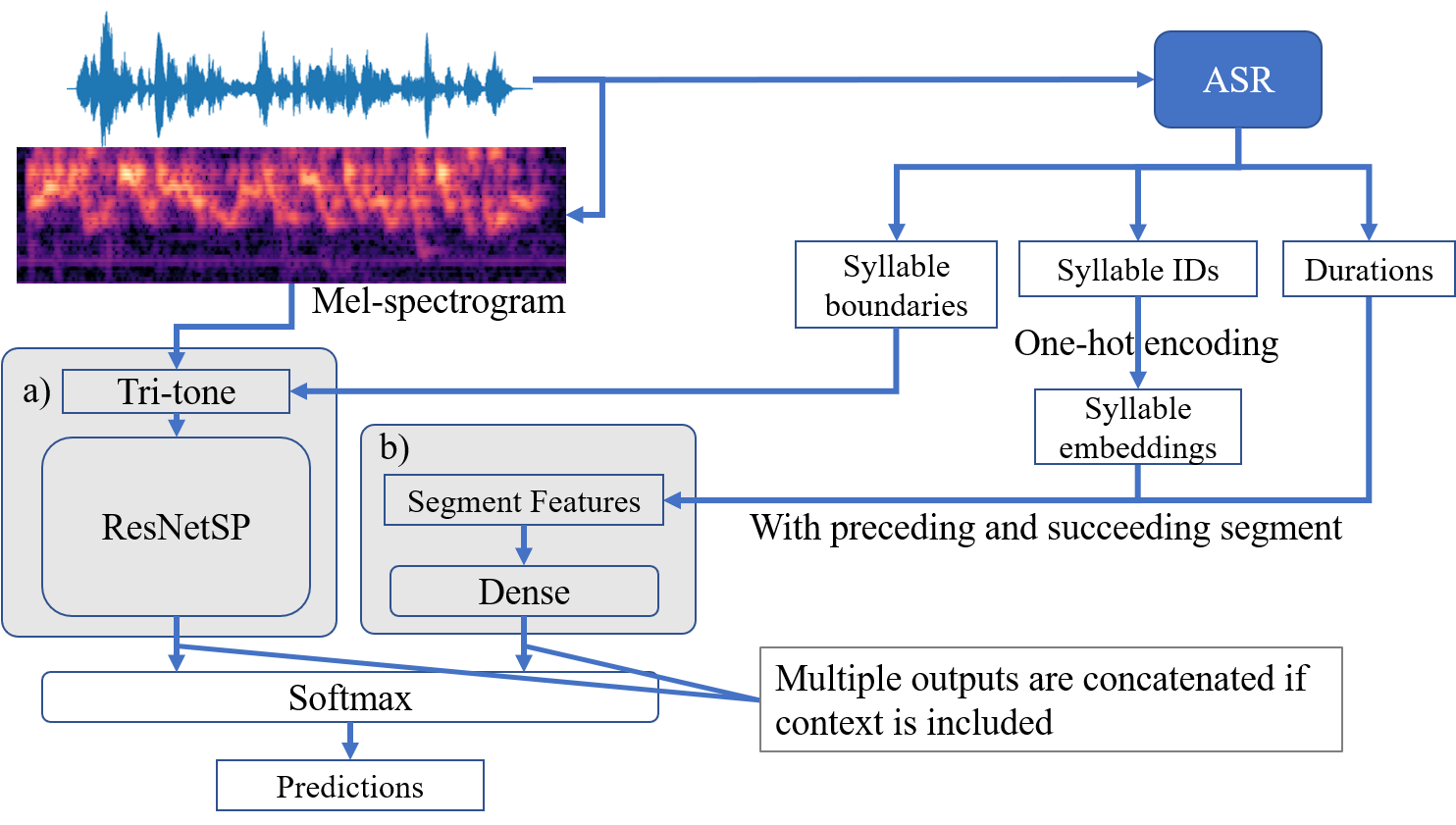}
    \end{center}
    \caption{Overview of the proposed framework}
    \label{fig:overall}
  \end{figure*}

  \subsection{Related Works}

  In previous literature, there are two main approaches of tone classification, namely the segment-based
  approach and the frame-based approach.

  In the segment-based approach, speech signal frames are cut into segments representing each syllable in a sentence.
  Then segment features are fed to the classifier to produce the tone label of the segments.
  There are many ways of segmenting the frames, such as manual annotation \cite{ToneNet},
  using force alignment \cite{tone_recognition_without_pitch,mandarin_tone_modeling_rnn,ditone},
  using the tone nucleus model \cite{tone_nucleus}, or using unsupervised methods \cite{tone_classification_cnn}.
  Also, the context information can be extracted from each segment and used to boost the performance of the model as
  well \cite{tone_recognition_without_pitch,mandarin_tone_modeling_rnn,tone_gabor}.
  Because the segment-based model relies heavily on force alignment, there might be a performance drop if
  the force alignment produces inaccurate results.
  In addition, this approach requires two passes, so the training process is usually more complex than a frame-based approach.

  In the frame-based approach, a sequence of frames is directly fed to the classifier,
  which outputs a sequence of labels.
  Techniques such as pooling \cite{tone_gabor}, attention \cite{tone_recognition_conv_blstm_att},
  or Connectionist Temporal Classification (CTC) \cite{tone_recognition_ctc} are often used
  to either group the frame-level output into segments or align the frames with phone boundaries
  so that the model can output a series of tone labels.
  To capture context information, layers like RNN and CNN can be used \cite{tone_recognition_conv_blstm_att,tone_recognition_ctc}.
  This approach uses a single pass to cover both the frame-segment conversion and the classification task and does not
  require a pre-trained ASR model.
  The downside of this approach is that the model is usually much more computational intensive due to the use of RNN,
  and it is difficult to evaluate the model on a per-tone basis, which is very important for mispronunciation
  detection.

  Besides, if the tone classifier is deployed in a CALL system, which naturally requires force alignment to calculate
  goodness-of-pronunciation (GOP) \cite{GOP}, a segment-based tone classification method can reuse the existing force
  alignment results, while the frame-based one cannot.

  For feature extraction, there are some popular choices in previous studies.
  Some researchers used spectrograms of audio signals in their experiments \cite{ToneNet,tone_gabor},
  while others used pitch tracking to extract the fundamental frequency contours of a speech utterance
  \cite{tone_nucleus,mandarin_tone_modeling_rnn,ditone}.
  MFCC is also a popular choice
  \cite{tone_classification_cnn,tone_recognition_conv_blstm_att,tone_recognition_without_pitch,tone_recognition_ctc}.

  \section{Proposed Framework}

  Our proposed framework uses a segment-based tone classification approach.

  The first part, called the embedding model, is a variant of Residual Neural Network (ResNet) \cite{resnet}
  which takes in a sequence of Mel-spectrograms and outputs tone embeddings.
  The second part, called the contextual model, is a multilayer perceptron model which takes in context segment
  features extracted from the force alignment results.
  The outputs of the two parts are concatenated and fed to the final classifier to produce the predictions.
  The overall architecture can be seen in Figure \ref{fig:overall}.
  The detailed configurations of the models will be explained in \ref{subsec:configurations}.

  \subsection{Feature Extraction of the Embedding Model}\label{subsec:feature-extraction-of-the-embedding-model}

  The Mel-spectrograms are extracted using the same method as ToneNet \cite{ToneNet}.
  More specifically, the waveforms are first resampled to $16000$ Hz, and the Mel-spectrograms are then extracted
  using $64$ Mel-filters, $[50, 350]$ Hz frequency range, and $13$ ms window length.

  Meanwhile, a pre-trained DNN-HMM ASR model is used to perform force alignment in order to obtain syllable boundaries
  and syllable IDs. Note that we cannot directly use this model as the baseline because this error rate takes insertion
  and deletion into account, while our model only produces one-to-one tone classification results.
  The syllable boundaries contain the start time and the duration of the syllables, either being an initial
  (\textit{shengmu}) or a final (\textit{yunmu}).
  The syllable IDs are unique identifiers of the syllables in Mandarin pronunciation, such as ``\textit{ai}'',
  ``\textit{h}'', ``\textit{ue}'', etc.
  Note that all tone information is removed from the syllables IDs used in our framework to
  prevent it from misleading our model.
  They are configured to be consecutive integers starting from $0$, as shown in Table \ref{tab:syllable_idx}.

  \begin{table}[h]
    \caption{Example syllable IDs (tone information removed)}
    \label{tab:syllable_idx}
    \centering
    \begin{tabular}{lllll}
      \hline
      sil & a & ai & an & \dots \\
      \hline
      0   & 1 & 2  & 3  & \dots \\
      \hline
    \end{tabular}
  \end{table}

  After that, the Mel-spectrograms are cut into syllable segments using the syllable boundaries.
  Since the syllable boundaries can have errors, we propose a segmentation method called tri-tone to improve
  the robustness of the boundaries.
  A tri-tone segment only includes half of the frames from the previous segment and half of the frames from the next segment.
  Models trained with tri-tone segments are more robust against tonal coarticulation because tri-tone
  segments include some information about the surrounding segments.
  In contrast to the di-tone segment \cite{ditone} which only contains frames in the next segments,
  our method preserves an additional section before the current segment,
  due to the bi-directional effect of tone coarticulation \cite{contextual_tonal_variations}.

\begin{figure}[h]
    \begin{center}
      \includegraphics[width=70mm]{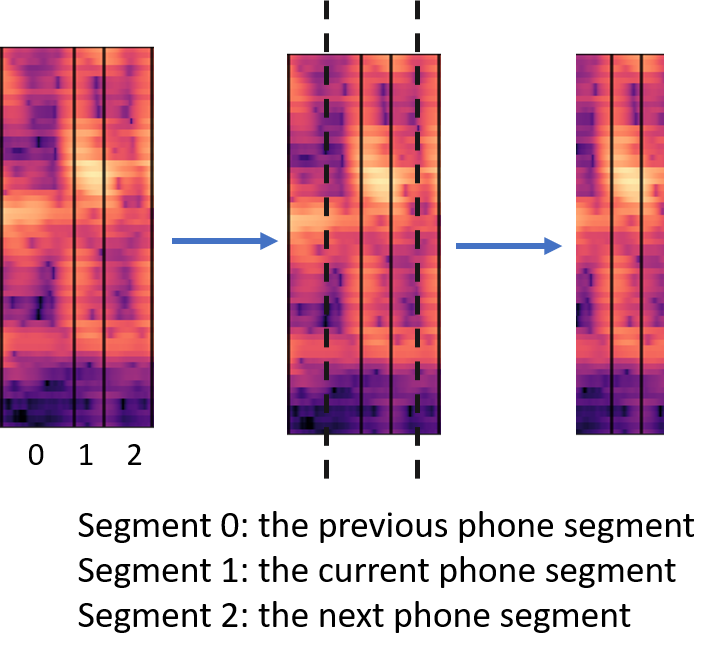}
    \end{center}
    \caption{Example of tri-tone segments being more robust when the force alignment results are inaccurate}
    \vspace*{-3pt}
    \label{fig:tritone}
  \end{figure}

  Figure \ref{fig:tritone} illustrates how tri-tone segments are more tolerant to inaccurate force alignment (dashed
  line) than syllable segments (solid line).

\begin{figure}[h]
    \begin{center}
      \includegraphics[width=40mm]{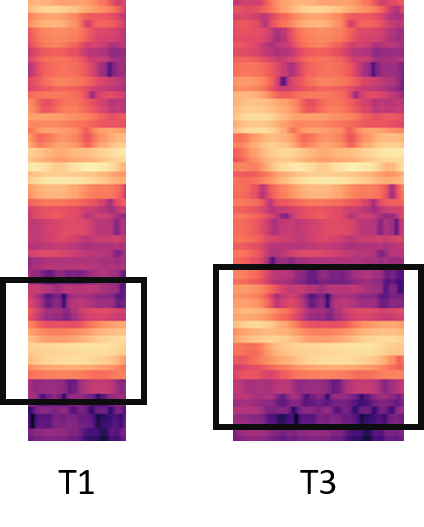}
    \end{center}
    \caption{An example of frame segmentation that causes ambiguities because it includes too many frames in neighbor
    segments}
    \label{fig:ambiguous_segment}
  \end{figure}

  The purpose of including only half of the frames from neighboring segments is to avoid ambiguities in terms of the overall pitch curve.
  For example, the spectrogram on the right part of Figure \ref{fig:ambiguous_segment} includes all frames in neighboring
  segments, where the pitch contours suggest T3, while in reality, it is T1.
  Meanwhile, a tri-tone segment accurately reflects a T1, shown on the left part of the figure.

  In addition, segment features are extracted from force alignment results.
  A segment feature consists of the duration and the syllable embedding of the current syllable.
  The duration has already been proved to be useful in previous literature
  \cite{tone_nucleus,tone_recognition_without_pitch,mandarin_tone_modeling_rnn}.
  Since different finals can have the same tone in Mandarin,
  adapting the model to different types of finals can boost the model's performance.
  Similar to speaker adaptation methods \cite{ivector_spk_adapt}, we can give the model a vector that uniquely identifies individual syllables,
  and then the tone classifier can adapt to different syllables to have a better performance.
  Since the alignment results from the ASR model are in syllable-level transcripts, we can reuse this
  information to create syllable embeddings, which is a one-hot encoding of the syllable index, shown in Table \ref{tab:syllable_idx}.

  Note that this method assumes that the annotated syllable IDs match the actual pronunciation if there are any pronunciation
  errors made by the speakers.
  If the syllable embeddings do not indicate the actual syllable pronunciation, the syllable embeddings provide false information to
  the model and lead to performance degradation
  However, we do not need to consider this case because all of AISHELL-3's labels represent the actual pronunciation.

  \subsection{Feature Extraction of the Contextual Model}

  Since context information is important for continuous tone recognition, we propose contextual features to cover such information.
  As the input of the contextual model, the contextual features contain 1) context tri-tone and 2) context segment features.

  A tri-tone context consists of the tri-tone segments of $n$ preceding, current, and $n$ succeeding syllables, where $n$ is the context size.
  Similarly, the context segment feature includes the segment features of $n$ preceding, current and succeeding syllables.

  Note that the tri-tone segment and the contextual features are different concepts.
  A tri-tone segment is not contextual as it is the feature of one syllable with soft boundaries,
  while the contextual features hold the information of multiple syllables.

  \section{Data}

  AISHELL-3 \cite{aishell3} is used in the following experiments.
  AISHELL-3 is a large-scale Mandarin speech corpus with Chinese pinyin transcripts.
  It contains approximately $85$ hours of mono-channel recordings from $218$ native Mandarin speakers with a sample rate of
  $44.1$ kHz.

  The transcripts are labeled manually to reflect the actual pronunciation perceived by a human
  and to overcome the difficulties caused by Homography, Tone Sandhi, Erization, and more \cite{aishell3}.

  For our experiments, the data is split into a training set and a testing set with $174$ and $44$ speakers, respectively.
  The train set contains $65.6$ hours of speech, and the test set contains $9.4$ hours of speech.
  There is no speaker overlap between the two subsets.

  \begin{table}[h]
    \caption{Number of samples of each tone in training data}
    \label{tab:train_samples_per_tone}
    \centering
    \begin{tabular}{lllllll}
      \hline
      Tone       & T0     & T1     & T2     & T3     & T4     & T5    \\
      \hline
      Count      & 751150 & 160885 & 179606 & 122707 & 253441 & 46609 \\
      Percentage & 49.6   & 10.6   & 11.9   & 8.10   & 16.7   & 3.08  \\
      \hline
    \end{tabular}
  \end{table}

  The number of data samples in the training set per tone is shown in Table \ref{tab:train_samples_per_tone}.
  Since there is a significantly larger amount of initials in the training data than other tones, a portion of initials
  is randomly removed during the training process to avoid performance degradation caused by data imbalance.
  The probability of removing an initial is equal to the inverse of the proportion of the number of initials
  to the total number of samples in the training set.

  \section{Experiments}

  \begin{table*}[t]
    \caption{Overall accuracy, the f1-scores per tone, and accuracy on tone patterns}
    \label{tab:results}
    \centering
    \begin{tabular}{llllllll}
      \hline
      & \textbf{ResNetSP} & \textbf{+SF} & \textbf{+SF+C1} & \textbf{+SF+C2} & \textbf{+SF+C3} & \textbf{+SF+C4} & \textbf{+SF+C5} \\
      \hline
      Overall Accuracy & 0.795             & 0.847        & 0.898           & 0.907           & 0.924           & 0.925           & \textbf{0.926}  \\
      T0               & 0.939             & 0.999        & 0.999           & 0.999           & 0.999           & 0.999           & 0.999           \\
      T1               & 0.840             & 0.856        & 0.903           & 0.910           & 0.925           & 0.927           & \textbf{0.929}  \\
      T2               & 0.760             & 0.802        & 0.867           & 0.874           & 0.894           & \textbf{0.897}  & 0.896           \\
      T3               & 0.729             & 0.781        & 0.856           & 0.874           & \textbf{0.902}  & 0.901           & \textbf{0.902}  \\
      T4               & 0.777             & 0.805        & 0.854           & 0.875           & 0.899           & 0.899           & \textbf{0.904}  \\
      T5               & 0.638             & 0.820        & 0.914           & 0.916           & 0.928           & \textbf{0.930}  & 0.929           \\
      \hline
      T4-T4            & 0.690             & 0.719        & 0.817           & 0.830           & 0.879           & 0.886           & \textbf{0.893}  \\
      T4-T3-T4         & 0.763             & 0.807        & 0.891           & 0.889           & 0.918           & \textbf{0.928}  & 0.923           \\
      T2-T2            & 0.717             & 0.761        & 0.855           & 0.842           & 0.864           & 0.862           & \textbf{0.876}  \\
      \hline
    \end{tabular}
  \end{table*}

  \subsection{Data Processing}

  A time delay neural network (TDNN) \cite{tdnn} model \cite{openslr_m10} pre-trained using the data from
  DataTang \cite{datatang1505hours} is used to perform force alignments.
  The model is deliberately chosen to have a different training dataset than the one used in our experiment to
  train the tone classifier in order to include imperfect force alignment results.
  The syllable error rate of this model tested on AISHELL-3 is $6.29\%$.

  Then we obtain the tri-tone Mel-spectrogram segments as describe above.
  When the tri-tone segments are fed to the neural network in mini-batches, samples are zero-padded to the maximum
  length found in the batch.

  \subsection{Configurations}\label{subsec:configurations}

  In all of the following experiments, a variant of ResNet is used, shown in Figure \ref{fig:overall}(a).
  This model is the same as the one used for far-field speaker recognition \cite{resnet34statspool},
  except the embedding layer contains only 128 units.
  For convenience's sake, this model will be referred to as ResNetSP.

  Cross entropy is used as the loss function,
  and Stochastic Gradient Descent (SGD) with $0.001$ as the learning rate
  and $0.9$ as the momentum is used as the optimizer.
  Also, early stopping and a ReduceOnPlateau learning rate scheduler are used.

  In the first experiment, the ResNetSP model is trained without any segment features or context information.
  This experiment is referred to as \textbf{ResNetSP} in the results.

  The second experiment is to train a classifier with tone embedding and segment features without context, using the
  architecture shown in Figure \ref{fig:overall}(a) plus Figure \ref{fig:overall}(b).
  In this configuration, the segment features are passed to a dense layer first, and the output of ResNetSP is
  concatenated to the output of the dense layer.
  The concatenated vector is fed to a softmax output layer, which has 128 units.
  This experiment is referred to as \textbf{+SF} in the results.

  The third experiment is performed in the same way, except the context size is set to 1.
  Compared to the previous configuration, all tri-tone segment features in a context are fed to the ResNetSP model one at
  a time, and the outputs of the segments are concatenated into a context tone embedding.
  Moreover, the context segment features are used, so the weight matrix of the dense layer mentioned in the second
  experiment is larger.
  This experiment is referred to as \textbf{+SF+C1} in the results.

  Other experiments are performed with different context sizes.
  They are referred to as \textbf{+SF+C}$n$ in the results, where $n$ is the context size.

  \section{Results and Discussions}

  The results of the experiments are shown in Table \ref{tab:results}.

  In the first experiment, ResNetSP achieves an overall accuracy of $79.5\%$.
  Meanwhile, the f1-scores of T3 and T5 are the lowest among all tones.
  This matches the results in previous studies
  \cite{tone_nucleus,tone_recognition_conv_blstm_att,tone_recognition_without_pitch,ditone}.
  Note that the f1-score of T0 is very high.
  This is likely because the model can easily recognize the fact that there is usually a fast pitch transition before
  and a smooth pitch transition after an initial.

  Starting from here, we use \textbf{ResNetSP} as a baseline and compare our proposed framework to it.
  As discussed in previous sections, we expect that by adding segment features and context information, the model will
  have a better performance.
  In particular, i) the f1-score of T5 should be boosted significantly by segment features,
  and ii) by adding context information, tonal coarticulation should have a much less negative impact on
  the performance.

  Indeed, the results match our expectations.
  The overall accuracy of \textbf{+SF} and \textbf{+SF+C1} increase to $84.7\%$ and $89.8\%$, and f1-scores per-tone
  also increase significantly.
  Note that by adding segment features, the f1-score of T0 becomes nearly one.
  The reason is that syllable IDs themselves can be used to distinguish initials from finals.
  Additionally, even without the context features, the f1-score of T3 rises by a large margin, as shown in \textbf{+SF}
  and \textbf{ResNetSP}.
  This is likely because T3 has the longest average duration among all tones \cite{contextual_tonal_variations}.
  Finally, the difference between the f1-scores of T3 and other tones became much less when context
  information was added.
  This suggests the importance of context information in helping the model learn about tonal coarticulation.

  In \textbf{+SF+C4} and \textbf{+SF+C5}, there is no significant performance gain or degradation brought by
  introducing a larger context size.
  This suggests that the syllables far away from the current syllable do not contribute much to its pitch
  pattern formation, which matches conclusions from previous prosodic studies that the coarticulation effect
  caused by neighboring syllables diminishes over time \cite{contextual_tonal_variations}.
  Therefore, a long-term context is not very helpful in terms of performance,
  compared to a short-term one.

  On top of the overall accuracy and the per-tone f1-scores, some common tone subsequences prone to tonal coarticulation
  in sentences are selected to test the model.
  We call these tone subsequences tone patterns for convenience's sake.
  Three sequences prone to tonal coarticulation, T4-T4, T4-T3-T4, and T2-T2, are selected based on previous
  studies \cite{tonal_variation_analysis,tone_coart_in_tianjin} and our data analysis.
  The test accuracy of individual tones in the sequences is shown in the bottom half of Table \ref{tab:results}.

  As expected, as the context size increases, the accuracy of the tone patterns generally increases,
  suggesting again that the model can handle tonal coarticulation better using segment features and context
  information.
  However, \textbf{+SF+C5} does not achieve a higher performance in one of the tone patterns,
  compared to \textbf{+SF+C4}.
  This confirms that a long-term context does not provide too much useful information.

  In summary, our experiment results show that i) segment features are useful, especially for T0 and T5,
  ii) combined with the context information, the model can deal with tonal coarticulation better,
  and iii) adding a long-term context does not boost the model's performance by much.

  \section{Conclusions and Future Work}

  In the proposed tone classification framework, by reusing the existing information from the force alignment, segment
  features and context information are extracted to improve the performance of tone classifiers by enhancing the
  robustness against tonal coarticulation.
  In addition, experiment results show that a short-term context is preferred rather than a long-term one.
  However, there are some possible improvements to this study.
  First, the model is both trained and tested on a dataset containing only native speakers.
  However, to evaluate the performance in terms of CALL applications, the model should be tested on an L2 dataset.
  Second, our work can be improved by performing data augmentation, such as adding slight noise and applying speed perturbation.
  The imbalanced data issue can also be handled better by data augmentation instead of random downsampling.
  Finally, we should perform experiments using RNN or Transformer models to verify our conclusion about the effect of
  long-term context.

  \section{Acknowledgements}

  This work is supported by the Teaching and Assessment Grant of Duke Kunshan University.

  \bibliographystyle{IEEEtran}

  \bibliography{mybib}

\end{document}